\documentclass[aps,pra, twocolumn, nofootinbib, showpacs]{revtex4-1} 
\usepackage[unicode=true,pdfusetitle,bookmarks=true,bookmarksnumbered=false,bookmarksopen=false,breaklinks=false,pdfborder={0 0 0},backref=false,colorlinks=false] {hyperref}
\hypersetup{colorlinks,linkcolor=myurlcolor,citecolor=myurlcolor,urlcolor=myurlcolor}
\usepackage{braket,colortbl,amsthm,amsmath,cleveref,amssymb,txfonts}\definecolor{myurlcolor}{rgb}{0,0,0.7}
\usepackage{bbm}
\usepackage{graphics,graphicx}
\usepackage{color}
\usepackage{amssymb}
\usepackage{amsthm}
\usepackage{amsfonts}
\usepackage{float}
\usepackage{graphicx}
\usepackage[format = plain,labelfont = bf,up, textfont = normal , up, justification=raggedright, singlelinecheck =false]{caption}
\usepackage{subcaption}
\usepackage{tabularx}
\usepackage{amsmath}
\usepackage{braket}
\DeclareMathOperator{\tr}{Tr}
\usepackage{graphicx}
\usepackage[utf8x]{inputenc}
\usepackage{color,soul}
\usepackage{amsmath}
\usepackage{braket}
\usepackage{latexsym}
\usepackage{amssymb}
\usepackage{amsthm}
\usepackage{xcolor}
\usepackage{bm}
\usepackage{graphics,epstopdf}
\usepackage{color}\usepackage{amsmath}
\usepackage{enumitem}
\usepackage{fmtcount}
\usepackage{booktabs}
\usepackage{appendix}
\usepackage{cancel}
\usepackage{epsfig}
\def\tr{\text{tr}}
\theoremstyle{plain}

\begin{document}
\title{Incompatibility of local measurements provide advantage in local quantum state discrimination}

\author{Kornikar Sen}
\affiliation{Harish-Chandra Research Institute,  A CI of Homi Bhabha National Institute, Chhatnag Road, Jhunsi, Allahabad 211 019, India}

\author{Saronath Halder}
\affiliation{Harish-Chandra Research Institute,  A CI of Homi Bhabha National Institute, Chhatnag Road, Jhunsi, Allahabad 211 019, India}

\author{Ujjwal Sen}
\affiliation{Harish-Chandra Research Institute,  A CI of Homi Bhabha National Institute, Chhatnag Road, Jhunsi, Allahabad 211 019, India}


\begin{abstract}
The uncertainty principle may be considered as giving rise to the notion of incompatibility of observables, a property that has been carefully analyzed in the literature for single systems. A pack of quantum measurements that cannot be measured simultaneously is said to form a set of incompatible measurements. Every set of incompatible measurements have advantage over the compatible ones in a quantum state discrimination task where one prepares a state from an ensemble and sends it to another party, and the latter tries to detect the state using available measurements. Comparison between global and local quantum state discriminations is known to lead to a phenomenon of ``nonlocality". In this work, we seal a connection between the domains of local quantum state discrimination and incompatible quantum measurements. We consider the local quantum state discrimination task where a sender prepares a bipartite state and sends the subsystems to two receivers. The receivers try to detect the sent state using locally incompatible measurements. We analyze the ratio of the probability of successfully guessing the state using incompatible measurements and the maximum probability of successfully guessing the state using compatible measurements. We find that this ratio is upper bounded by a simple function of robustnesses of incompatibilities of the local measurements. 
Interestingly, corresponding to every pair of sets of incompatible measurements, there exists at least one local state discrimination task where this bound can be achieved. We argue that the optimal local quantum state discrimination task does not present any ``nonlocality'', where the term is used in the sense of a difference between the ratios, of probabilities of successful detection via incompatible and compatible measurements,  
in global and local state discriminations. The results can be generalized to the regime of  multipartite local quantum state distinguishing tasks.
\end{abstract}
\maketitle

\section{Introduction}
The uncertainty principle is one of the fundamental pillars that influenced the formation of quantum mechanics by introducing us to the concept of incompatibility of observables.
Given two observables, if the operators corresponding to those observables - within the quantum formalism - can be jointly measured using a ``parent'' measurement,
then the observables are called ``compatible''. Otherwise, they are ``incompatible''~\cite{inc0.5, inc1,inc22,inc3,inc4,inc5,inc6,inc7}. Incompatibility is entirely a single-system property, i.e., a system considered as a whole even if it possesses multiple constituents. Incompatibility of observables is a  signature quantum mechanical property, absent in classical systems, and plays an important role in many quantum  tasks and phenomena, like quantum key distribution \cite{qkd1, qkd2, qkd3, qkd4}, quantum steering \cite{qs1, qs2, qs3, qs4, inc7, inc3}, etc. 

In a few recent works, connections between  minimum-error quantum state discrimination and  incompatibility of measurements have been explored~\cite{toigo1, cavalcanti1, Guhne, inc5, inc1, inc2}. The state discrimination task involves a sender, Alice, and a receiver, Bob. 
Alice prepares a quantum system in a particular state, taken from a particular ensemble 
and then sends the system to 
Bob, who is possibly  at a distance. The ensemble 
appears at Alice with a certain given probability, and
along with the quantum system, Alice may also send the information regarding the ensemble
to Bob. The set of possible ensembles and their constituents are known to both the parties. After receiving the system, Bob tries to identify the state of the system through measurements, i.e., Bob tries to distinguish between the states of the ensemble.
It is possible that Bob has access to a fixed set of measurements. Depending on the available set of measurements, 
Bob may not be able to 
identify the state of the system perfectly. In such a state discrimination problem, 
Bob can try to identify the state of the system through what is 
known as the minimum-error quantum state discrimination strategy~\cite{mes1,mes2, mes3, mes4}, by minimizing
the overall probability of error in guessing the state of the system.

In Ref.~\cite{toigo1}, the authors have considered a particular type of state discrimination task where the sender may provide some information about the state before the receiver performs any measurement. The optimal guessing probability when the pre-measurement information is provided is equal to the optimal guessing probability when the information is given after the measurement if the available set of measurements are compatible. This implies that pre-measurement information can improve the situation if the measurements are incompatible. The maximum advantage one can get from incompatibility increases linearly with the robustness of incompatibility~\cite{cavalcanti1}. There exists certain state discrimination tasks where incompatibility provides advantage, and thus incompatibility of measurements can be regarded as a resource~\cite{Guhne}. The collection of the compatible set of measurements forms a closed and convex set, and in Ref.~\cite{inc5}, a witness operator is formulated to detect incompatible measurements. We 
mention here that 
a relation between quantum state discrimination and channel incompatibility has also been established~\cite{inc_chan1, inc_chan2}.
Hitherto, in research works where incompatible measurements are examined in the context of their ability to discriminate quantum states, a global state discrimination task has been considered. The receiver is allowed to do measurements on the entire state considering it as a single entity. We want to explore the situation where the sender sends each part of the system to different receivers, so that the receivers, situating at distant locations, are not able to perform measurement on the entire system but are only allowed to do local measurements.

There exists unique and interesting properties of distributed quantum systems which can provide advantages in many quantum devices over the corresponding classical ones. The difference between the ability to distinguish shared quantum states using global and local operations provides evidence of “nonlocality” present in the considered situation, which itself is an interesting quantum phenomenon, but is also of crucial importance in several quantum tasks.

\begin{figure}
\includegraphics[scale=0.47]{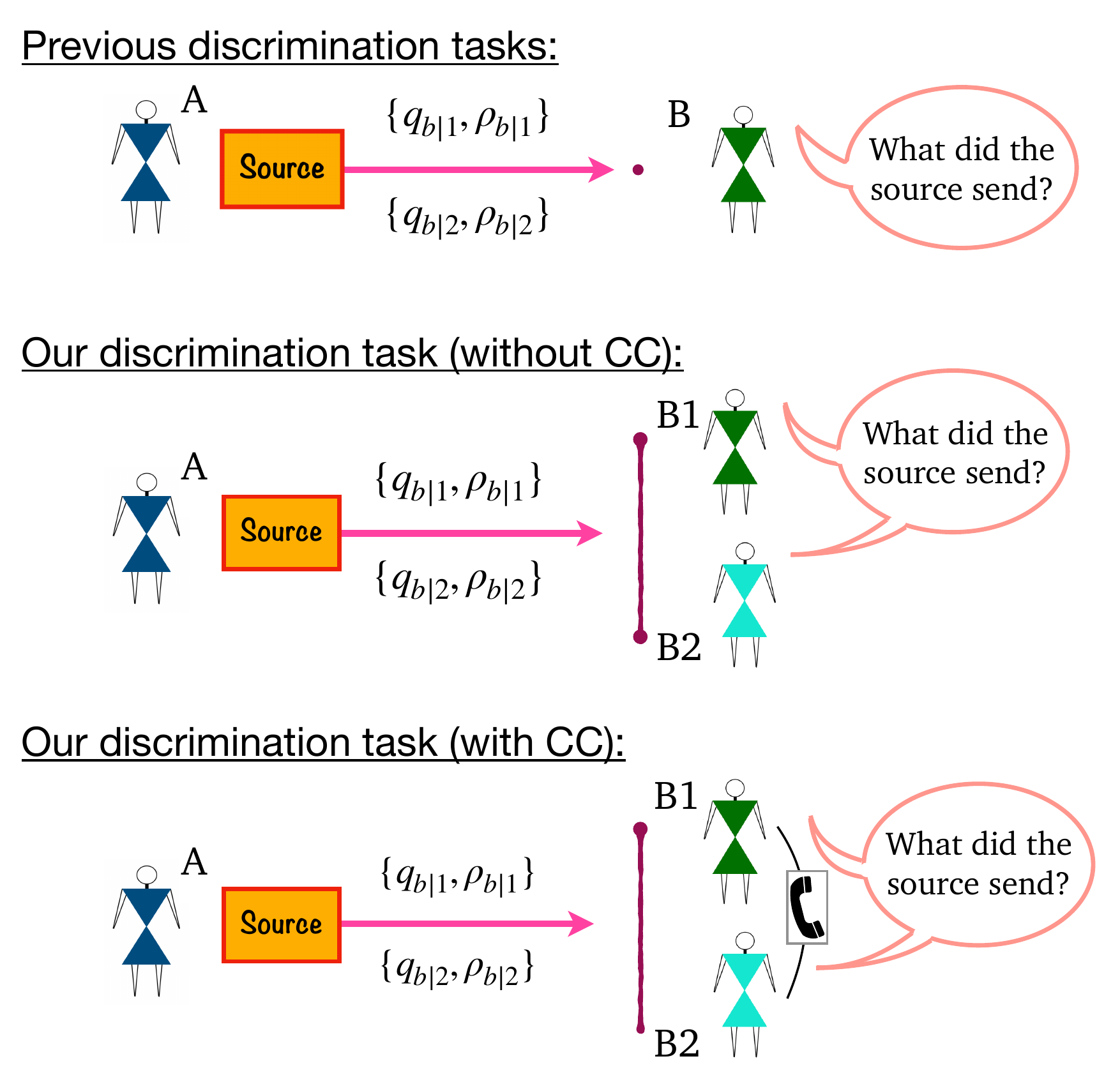}	\caption{Comparison between the discrimination task considered in previous literature and the one explored in this paper, with respect to incompatible measurements. Incompatible measurements are known to provide advantage over compatible ones in certain quantum state discrimination tasks. We have schematically depicted such a discrimination task where a girl, say Alice, randomly chooses a state from a randomly selected ensemble (among a given set of ensembles) and sends it to a boy, Bob (see upper panel). Bob has access to a set of measurements using which he tries to distinguish the state. In our protocol, Alice again randomly selects a state from a random ensemble with the only difference that in this case, the ensembles consist of bipartite states. Alice sends each party of the bipartite state to distant locations, say to Bob1 and Bob2. Bob1 and Bob2 being located at two different places are unable to perform any joint operation on the entire state consisting two parts. Thus Bob1 and Bob2 can either do only local operations without any classical communication (depicted in middle panel) or local operation along with classical communication (depicted in lower panel) on their part of the system. Whatever be the operations, be it global or local, the aim of the receivers, i.e., Bob or Bob1 and Bob2 is to distinguish the received state. For more details see the main text.}
	\label{fig1}
\end{figure}

In this work, we try to combine these two fundamental notions of quantum mechanics, viz. incompatibility of quantum measurements and nonlocality in state discrimination of shared quantum systems. Specifically, we want to examine if the single-system property of measurement incompatibility can influence the quantum state discrimination protocol of a shared system.
Therefore, we consider quantum state discrimination tasks where more than two parties are involved. Precisely, a sender, Alice, prepares a 
quantum system of \emph{more than one subsystem} in a particular state and then sends the subsystems to two or more spatially separated parties. These parties try to identify the state of the system but they are not allowed to 
employ
quantum communication between the spatially separated locations. In this situation, the allowed class of operations can be categorized into two groups, depending on the resources available: (i)~local quantum operations~(LO) without classical communication~\cite{LO1, LO2, LO3}  and (ii)~local quantum operations and classical communication~(LOCC)~\cite{LOCC3.5, LOCC3, LOCC4, LOCC5, LOCC6, LOCC7, LOCC9, LOCC10, LOCC11, LOCC12, LOCC13, LOCC14, LOCC15, LOCC16, LOCC17, LOCC18, LOCC19, LOCC20, LOCC21, LOCC22, LOCC24, LOCC25, LOCC26, LOCC27, LOCC28, LOCC29, LOCC30, LOCC31, LOCC32, LOCC34, LOCC33, LOCC35, LOCC36, LOCC38, LOCC40, LOCC44, LOCC45, LOCC49, LOCC50, LOCC55, LOCC56, LOCC65, LOCC66, LOCC67, LOCC68}. In Fig.~\ref{fig1}, we schematically present a comparison between the state discrimination tasks considered in previous literature in the context of determining advantage of incompatible measurements with our discrimination task.

\begin{figure*}
\includegraphics[scale=0.27]{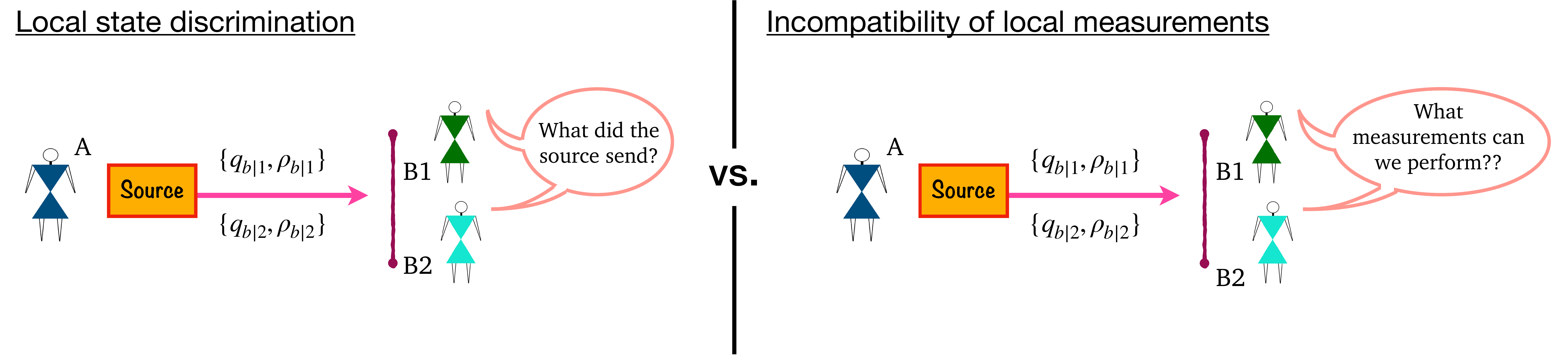}	\caption{Schematic presentation of the two physical phenomena that we want to relate in this paper. In the left panel, we show the local state discrimination task, where Alice sends each part of a randomly chosen bipartite state from a randomly selected ensemble (among a given set of ensembles) to Bob1 and Bob2. The receivers, Bob1 and Bob2, using local operations, want to discriminate the state. In the right panel, the situation is the same, but here Bob1 and Bob2 are deciding which measurements should be performed to maximize the probability of successful discrimination. Should the measurement be chosen from an incompatible set of measurements, or would a compatible set provide more advantage? Here, we have not considered classical communication. Similar figures can be drawn considering classical communication as well.}
	\label{fig2}
\end{figure*}

We establish connections for the 
two categories of \emph{local} state discrimination tasks
with the incompatibility of available local measurements. See Fig.~\ref{fig2} to get a schematic understanding of the two phenomena which we are trying to bring in the same context.
The spatially separated parties have access to 
sets of incompatible measurements, employing which they try to accomplish the given state discrimination task. We derive relations between the probability of successfully guessing (PSG) the state of the system using local incompatible measurements and  incompatibility of the local measurements. We provide  upper bounds, considering LO and LOCC separately, and analyzing  a single round of measurements in the latter case, on the PSG and these upper bounds are the functions of incompatibility of local measurements. 
Interestingly, corresponding to every set of incompatible local measurements there exists at least one local state discrimination task in which this upper bound can be reached. 
The optimal state discrimination task which achieves this bound does not exhibit any ``nonlocality''.

	

\section{Incompatible measurements and robustness of incompatibility}.\label{sec2}
A set of measurements $\{ M_x\}_x$ is called compatible or incompatible depending on their joint measurability. The suffix, $x$, outside the braces in $\{M_x\}_x$ indicates the running variable that generates the set. Similar notation is used throughout the manuscript. If $\{ M_x\}_x$ can be measured simultaneously using a parent measurement, $G$, we say that it is  compatible. We denote the measurement operators, associated with different outcomes of a measurement $M_x$ and $G$, by $\{M_{a|x}\}_a$ and $\{G_\lambda\}_\lambda$ respectively. The measurement operators corresponding to the measurements $\{ M_x\}_x$ can be expressed in terms of $G$
as
%
\begin{equation}
M_{a|x}=\sum_{\lambda}p(a|x,\lambda)G_\lambda,
\end{equation}
where $p(a|x,\lambda)$ is a conditional probability distribution, $M_{a|x}, G_\lambda\geq0$, $\sum_a M_{a|x}=\mathbbm{1}$, and $\sum_\lambda G_\lambda=\mathbbm{I}$ $ \forall a,x,\lambda$  with $\mathbbm{1}$ being the identity operator.
 
Measurements which are not compatible (i.e., not jointly measurable) are called incompatible measurements. To quantify incompatibility of a set of measurements, $\{ M_k\}_k$, the robustness of incompatibility~(ROI), denoted $I_M$, was introduced in the literature (for example, see Ref.~\cite{cavalcanti1}). ROI can be defined by the minimal amount of {\it noise} that is required to be mixed with a set of incompatible measurement, $\{M_k\}_k$, to make it compatible, i.e., 
\begin{eqnarray}
I_M=\min r, && \label{sdp}\\
\text{such that}~~ &&  \frac{M_{c|k}+r\Lambda_{c|k}}{1+r}=\sum_{\lambda}p(c|k,\lambda)G_\lambda,\label{1}\\
&& \Lambda_{c|k}\geq 0\text{, }\sum_c \Lambda_{c|k}=\mathbbm{1},\label{2}\\
&& G_\lambda\geq 0\text{, }\sum_{\lambda}G_\lambda=\mathbbm{1},	\label{3}	\\
&& 0\leq p(c|k,\lambda)\leq 1\text{, }\sum_cp(c|k,\lambda)=1,\label{4} 
\end{eqnarray}
where $M_k = \{M_{c|k}\}_c$, i.e., $\{M_{c|k}\}_c$ are the outcomes of the measurement ${M_k}$. $I_M$ denotes the amount of incompatibility present in the set of measurements $\{M_k\}_k$. Eq.~\eqref{sdp} provides a generic definition for quantification of incompatibility. Precisely, the definition does not depend explicitly on the nature of $\{M_k\}_k$, i.e., if it is a set of projective measurements (PM) or positive operator valued measurement (POVM). Thus each element of the set $\{M_k\}_k$ just satisfies the usual properties of a measurement, $M_k\geq 0$ and $\sum_c M_{c|k}=\mathbb{I}$. By {\it noise}, we  mean an arbitrary set of measurements $\{\Lambda_k\}_k$, $\Lambda_k = \{\Lambda_{c|k}\}_c$, which is  mixed with 
the set of measurements $\{ M_k\}_k$ so that after mixing, the final set of measurements, $\frac{M_{c|k}+r\Lambda_{c|k}}{1+r}$, become compatible. $r$ is the amount of noise that has to be mixed with $\{ M_k\}_k$ to make the final measurement compatible. The minimization is over $r$, $\Lambda_{c|k}$, $G_\lambda$, and probability distributions, $p(c|k,\lambda)$. By minimization over conditional probability distributions we mean minimization over any set of real numbers, $\{p(c|k,\lambda)\}$, which satisfies $p(c|k,\lambda)\geq 0$ for all $c$, $k$, and $\lambda$, and $\sum_{c} p(c|k,\lambda)=1$ for all $k$ and $\lambda$. Whenever we do any optimization over conditional probabilities, we use this same concept. $\{\Lambda_k\}_k$ and $\{G\}$ represent measurements with outcomes $\{\Lambda_{c|k}\}_c$ and $\{G_\lambda\}_\lambda$, respectively. These measurements can also be POVM as well as PM. The constraints on these measurement operators are mentioned in the expressions~(\ref{1}-\ref{4}). 

\section{Connection of incompatibility with  state discrimination problem}.\label{sec3}
In several articles \cite{toigo1,inc5,cavalcanti1,Guhne}, it has been shown that incompatible measurement can provide advantage over the compatible ones in certain state discrimination tasks. In Ref~\cite{cavalcanti1}, the authors have considered a task involving two people where one randomly selects a state from an ensemble, $\mathcal{E}_y$, which has been randomly chosen from a given set of ensembles, $\{\mathcal{E}_y\}_y$, and sends the state to the other. The latter, after receiving the state, tries to discriminate it using a set of measurements. Let $\mathcal{P}^C(\{\mathcal{E}_y\})$ and $\mathcal{P}^I(\{\mathcal{E}_y\},\{Q_x\})$ be the maximum probability of successfully guessing the state maximized over all set of compatible measurements and the same for a fixed set of measurements, $\{Q_x\}$, for the optimal strategy. In Ref.~\cite{cavalcanti1}, a precise mathematical expression is provided  to represent the advantage achievable through incompatible measurements which goes as follows: 
\begin{equation}
 \frac{\mathcal{P}^I(\{\mathcal{E}_y\},\{Q_x\})}{\mathcal{P}^C(\{\mathcal{E}_y\})}\leq 1+I_Q.\nonumber
\end{equation}
Here $I_Q$ is the ROI of $\{Q_x\}$. The authors have also shown that for any set of measurements there exists a corresponding state discrimination task for which the bound is achievable.

In this work, we restrict ourselves to a smaller set of operations, i.e., we consider state discrimination using only local operations with or without classical communication instead of global measurements, and try to determine how the above expression gets modified in the new situation. The considered state discrimination tasks are discussed in detail below.

\emph{State discrimination using only LO without CC.}--If the spatially separated parties are restricted to perform local operations only, on their subsystems, in order to accomplish the given state discrimination task, then it is known as state discrimination by LO. We note that in this scenario, classical communication among the parties during the local operations within a round or between the rounds, is not allowed. However, after all the local operations are accomplished, the parties can discuss the measurement outcomes to identify the state of the system \cite{LO1, LO2, LO3}. Since in this scenario classical communication is not allowed in between the measurements, we will denote the corresponding probability of successful guess using the suffix LO\cancel{CC}.


\emph{State discrimination task with pre-measurement information}.--In this task, Alice chooses an ensemble, $\mathcal{E}_y$, of \textit{bipartite states} with probability $q(y)$. Then she prepares a quantum system in a bipartite state, $\rho_{b|y}$, taken from $\mathcal{E}_y$, with probability $q(b|y)$ and sends the subsystems to Bob1 and Bob2 along with the information of $y$. Bob1 and Bob2 have sets of measurements $\{M_k\}_k$ and $\{N_l\}_l$ respectively, using which Bob1 and Bob2 want to identify the state of the system. We refer to this as SD1. 
 
\emph{State discrimination task with post-measurement information}.--
 This task is the same as the preceding one except 
 that in this case, Bob1 and Bob2 do not have any information about $y$, prior to the measurement. After the performance of measurements, Alice informs them about the particular ensemble, $\{\mathcal{E}_y\}$, and then depending on $y$ and the measurement outcomes, Bob1 and Bob2 make a guess about $\rho_{b|y}$. This state discrimination task will be referred to as
 SD2.

Here we consider SD1,i.e., state discrimination with pre-measurement information and provide a relation between the probability of successfully guessing (PSG) the state of the system, using the local measurements $\{M_k\}_k$ and $\{N_l\}_l$, and the incompatibility of these measurements. We consider the case where only local operations (LO) are allowed. 

\section{Upper bound on the guessing probability using only local operations without classical communication}.
We consider here the case where Bob1 and Bob2 try to discriminate the state using LO without CC, and first the case where they have the knowledge of $y$ prior to the measurement. We restrict Bob1 and Bob2 from using classical communication. The set of measurements available to Bob1 and Bob2 are locally incompatible. After receiving the state $\rho_{b|y}$,  Bob1 (Bob2) chooses a measurement, $M_k$ ($N_l$), with probability $p(k|y)$ ($p(l|y)$). The probability of guessing the state correctly using these measurements is given by
\begin{equation}
\small
P^{\text{SD1}}_{\text{LO\cancel{CC}}} = \sum_{y,b,k,l,c,d} q(y)q(b|y)\tr[\rho_{b|y}M_{c|k}\otimes N_{d|l}]p(k|y)p(l|y)p(b|c,d,y), \label{eqq3}
\end{equation}
where $M_{c|k}$ and $N_{d|l}$ are the measurement operators, corresponding to the outcomes $c$, $d$, associated with the measurements $M_k$ and $N_l$ respectively. Note that, here state discrimination with pre-measurement information has been considered. After getting the outcomes $c$ and $d$, Bob1 and Bob2 call for a guess regarding the value of $b$, and this is according to the probability, $p(b|c, d, y)$. Then, the optimal PSG using measurement $M_k$ and $N_l$ is 

\begin{eqnarray}
\label{eqq4}
P_{\text{LO\cancel{CC}}}^I(\{\mathcal{E}_y\},\{M_k\},\{N_l\})=\max_{p(k|y),p(l|y),p(b|c,d,y)} \sum_{y,b,k,l,c,d} q(y)q(b|y)\nonumber\\
\tr[\rho_{b|y}M_{c|k}\otimes N_{d|l}]p(k|y)p(l|y)p(b|c,d,y).~~~~~
\end{eqnarray}
Here, the maximization over conditional probabilities represent optimization over any set of real numbers which satisfies the usual properties of a conditional probability distribution. For example, for the set of probabilities, $\{p(k|y)\}_k$, the conditions are $p(k|y)\geq 0$ for all $k$ and $y$ and $\sum_k p(k|y)=1$ for all $y$. The final set of conditional probabilities, $\{p^{max}(l|y), p^{max}(l|y), p^{max}(b|c,d,y)\}$, which maximize the function, can be used to strategize the task. Precisely, if Bob1 and Bob2 choose their measurement, $M_k$ and $N_l$ with probabilities $p^{max}(k|y)$ and $p^{max}(l|y)$ and after doing the measurement, if they guess about the state depending on the probability distribution $p^{max}(b|c,d,y)$, they will reach the maximum probability of success.
On the other hand, if Bob1 and Bob2 had locally compatible measurements, and had got the information of $y$ only after performing the measurement, then the PSG would have been
\begin{equation}
\small
P^{\text{SD2}}_{\text{LO\cancel{CC}}} = \sum_{y,b,k,l,c,d} q(y)q(b|y)\tr[\rho_{b|y}M_{c|k}\otimes N_{d|l}]p(k)p(l)p(b|c,d,y).\label{eqq5}
\end{equation}
Here, the suffix SD2 represents that state discrimination with post measurement information has been considered. Then the maximum PSG using locally compatible measurements in SD2, i.e., state discrimination with post-measurement information, can be written as
\begin{equation}
P^C_{\text{LO\cancel{CC}}}(\mathcal{E}_y)=\max_{{M_k},{N_l} \in CM, p(k),p(l),p(c,d,y)} P^{\text{SD2}}_{\text{LO\cancel{CC}}},\label{eqq6}
\end{equation}
where the maximization is taken over the set of locally compatible measurements, $CM$, and the probabilities, $p(k)$, $p(l)$, $p(c,d,y).$ 

Precisely, Eq. \eqref{eqq3} represents PSG of a shared state $\rho_{b|y}$ when the parties Bob1 and Bob2 knows about $y$ before the performance of any measurement. Thus they choose their measurements, respectively $M_{k}$ and $N_{l}$, randomly, following independent probability distributions, $p(k|y)$ and $p(l|y)$, where these probability distributions depend on $y$. In the next equation, i.e., Eq. \eqref{eqq4}, we optimize the probability presented in Eq. \eqref{eqq3}, $P^{\text{SD1}}_{\text{LO\cancel{CC}}}$, over all possible conditional probability distributions, $p(k|y)$, $p(l|y)$, and $p(b|c,d,y)$, to determine the best strategy for discrimination. In Eq. \eqref{eqq5}, we consider the case where Bob1 and Bob2 does not know about $y$ before the performance of the measurements. Here also they randomly choose the measurements, $M_{k}$ and $N_{l}$, but these random distributions, $p(k)$ and $p(l)$, do not depend on $y$. However, after doing the measurements, Alice tells Bob1 and Bob2 about $y$. Hence Bob1 and Bob2 can guess about $b$, depending on the information of $y$ as well as the measurement outputs $c$ and $d$. To guess the value of $b$ they follow the probability distribution $p(b|c,d,y)$. In the final equation, Eq. \eqref{eqq6}, we optimize $P^{\text{SD2}}_{\text{LO\cancel{CC}}}$, which is expressed in Eq. \eqref{eqq5} with respect to all possible strategies, $p(k)$, $p(l)$, $p(c,d,y)$, and all possible pairs of sets of measurements, $\{M_k\}_k$ and $\{N_l\}_l$ which are locally compatible.

From now on, whenever a set of locally incompatible measurements will be used for state discrimination, we will consider that $y$ is known to Bob1 and Bob2 prior to the measurement, and in state discrimination tasks using compatible measurements, we will assume pre-measurement information about $y$ is not available~\cite{toigo1,inc5,Guhne,cavalcanti1}.

Let  ROIs of the local measurements $\{M_k\}_k$ and $\{ N_l\}_l$ be $I_M$ and $I_N$ respectively. Moreover, let the optimization, shown in Eq.~\eqref{sdp}, be attained by  $\Lambda^*_{c|k}$, $p^*(c|k,\lambda)$, $G_\lambda^*$ for $\{M_k\}$ and $\Sigma^*_{d|l}$, $p^*(d|l,\lambda)$, $H_\lambda^*$ for $\{N_l\}$. Then, using Eq.~\eqref{sdp}, one can find:
\begin{eqnarray}
&&M_{c|k}\leq (1+I_M)\sum_{\lambda}p^*(c|k,\lambda)G_\lambda^*, \nonumber\\
\text{and}~~~~~&&N_{d|l}\leq (1+I_N)\sum_{\lambda}p^*(d|l,\lambda)H_\lambda^*. 
\end{eqnarray}
We take the tensor product of these  two inequations, to find:
\begin{eqnarray}\label{tensor}
M_{c|k}\otimes N_{d|l}\leq (1+I_M)(1+I_N)~~~~~~~~~~~~~~~~~~~~\\ \nonumber
\sum_{\lambda,\nu}p^*(c|k,\lambda)p^*(d|l,\nu)G_\lambda^* \otimes H_\nu^*. 
\end{eqnarray}
We now multiply both sides of the above inequality by $q(y)q(b|y)p(k|y)p(l|y)p(b|c,d,y)\rho_{b|y}$, and then sum over the parameters $c$, $d$, $k$, $l$, $b$, $y$. Thereafter, taking trace, it becomes:
\begin{widetext}
\begin{eqnarray}
\sum_{y,b,k,l,c,d} q(y)q(b|y)p(k|y)p(l|y)p(b|c,d,y)\tr[\rho_{b|y}M_{c|k}\otimes N_{d|l}]\leq(1+I_M)(1+I_N)\sum_{y,b,k,l,c,d,\lambda,\nu} q(y)q(b|y)p(k|y)p(l|y)p(b|c,d,y)\nonumber \\p^*(c|k,\lambda)p^*(d|l,\nu)\tr[\rho_{b|y}G^*_\lambda\otimes H^*_\nu].
\end{eqnarray}
We substitute $\sum_{k,l,c,d}p(k|y)p(l|y)p(b|c,d,y)p^*(c|k,\lambda)p^*(d|l,\nu)$ by a new conditional probability distribution, $p(b|\lambda,\nu, y)$. Thus we have
\begin{eqnarray}
\sum_{y,b,k,l,c,d} q(y)q(b|y)\tr[\rho_{b|y}M_{c|k}\otimes N_{d|l}]p(k|y)p(l|y)p(b|c,d,y)\leq (1+I_M)(1+I_N)\sum_{y,b,\lambda,\nu} q(y)q(b|y)p(b|\lambda,\nu,y)\tr[\rho_{b|y}G^*_\lambda\otimes H^*_\nu].
\end{eqnarray}
\end{widetext}
The expression $\sum_{y,b,\lambda,\nu} q(y)q(b|y)p(b|\lambda,\nu,y)\tr[\rho_{b|y}G^*_\lambda\otimes H^*_\nu]$ represents the PSG using a single pair of local measurements, viz., $\{G_\lambda\}_\lambda$ and $\{H_\nu\}_\nu$. Hence, it would be less than $P^C_{g,\text{LO\cancel{CC}}}(\{\mathcal{E}_y\})$ which is the PSG, optimized over all compatible measurements. So we can write
\begin{eqnarray}
\sum_{c,d,k,l,b,y} q(y)q(b|y)\tr[\rho_{b|y}M_{c|k}\otimes N_{d|l}]p(k|y)p(l|y)p(b|c,d,y)\nonumber\\
\leq (1+I_M)(1+I_N)P^C_{\text{LO\cancel{CC}}}(\{ \mathcal{E}_y\}).~~~~~~
\end{eqnarray} 
The above relation holds for all probability distributions $p(k|y)$, $p(l|y)$, and $p(b|c,d,y)$. Hence it also holds if we maximize the left hand side of the above inequality with respect to these probabilities. Therefore we get
\begin{equation}
P^I_{\text{LO\cancel{CC}}}(\{ \mathcal{E}_y\}, \{ M_k\},\{ N_l\})\leq (1+I_M)(1+I_N)P^C_{\text{LO\cancel{CC}}}(\{ \mathcal{E}_y\}),\nonumber
\end{equation}
so that 
\begin{equation}\label{ublo}\frac{P^I_{\text{LO\cancel{CC}}}(\{ \mathcal{E}_y\}, \{ M_k\}, \{N_l\})}{P^C_{\text{LO\cancel{CC}}}(\{ \mathcal{E}_y\})}\leq (1+I_M)(1+I_N).~~~~~~
\end{equation}
Note that the numerator and the denominator of Eq. \eqref{ublo} are for local operations without classical communication.

The same bound remains valid when Bob1 and Bob2 are allowed to use classical communication along with local operations. That is, if the maximum PSG using local operation and classical communication (LOCC) in presence of same set of incompatible measurements, $\{M_k\}$ and $\{N_l\}$ be $P^I_{\text{LOCC}}(\{ \mathcal{E}_y\}, \{ M_k\},\{ N_l\})$ and the maximum PSG using compatible measurements be $P^C_{\text{LO{CC}}}(\{ \mathcal{E}_y\})$, then
\begin{equation}
  \frac{P^I_{\text{LOCC}}(\{ \mathcal{E}_y\}, \{ M_k\},\{ N_l\})}{P^C_{\text{LOCC}}(\{ \mathcal{E}_y\})}\leq (1+I_M)(1+I_N),\label{ublocc}  
\end{equation}
See Appendix~\ref{A1} for a proof. It should be noted that the numerator and the denominator of Eq. \eqref{ublocc} are for local operations and classical communication.

Though the right hand side (RHS) of the inequalities, \eqref{ublo} and \eqref{ublocc}, are the same, the left hand sides (LHS) of the same, by definition, differ significantly. Precisely, LHS of \eqref{ublo} represents the ratio of PSGs in state discrimination using local operations without any classical communication, whereas LHS of \eqref{ublocc} describes the ratio of PSGs in state discrimination when classical communication is allowed along with local operations. Certainly, if we individually compare  the numerator and denominator of LHS of \eqref{ublo} with \eqref{ublocc}, we see $P^I_{\text{LO\cancel{CC}}}\leq P^I_{\text{LOCC}}$ and $P^C_{\text{LO\cancel{CC}}}\leq P^C_{\text{LOCC}}$, but interestingly, the ratios are found to be upper bounded by the same quantity, $(1+I_M)(1+I_N)$, which is the same as for LO without CC. 

Instead of this bipartite state discrimination task, we can also consider an $n$-partite state discrimination task, where Alice prepares an $n$-partite system and then sends the subsystems to Bob1, Bob2, etc. After receiving the subsystems, Bob1, Bob2, \(\ldots\), Bob$n$ tries to identify the state using a set of local measurements, say $\{O^1_{k_1}\}_{k_1}$, $\{O^2_{k_2}\}_{k_2}$, \(\ldots\), $\{O^n_{k_n}\}_{k_n}$ respectively. Let the ROI of $\{O^i_{k_i}\}_{k_i}$ be $I_i$. Using the same technique as described in the above, it is possible to show the following:
\begin{equation}
\frac{P^I_{\text{LO\cancel{CC}/LOCC}}\left(\{ \mathcal{E}_y\},\{O^1_{k_1}\},\{O^2_{k_2}\},\ldots\right)}{P^C_{\text{LO\cancel{CC}/LOCC}}(\{ \mathcal{E}_y\})}\leq \prod_{i=1}^n (1+I_i).\label{gen1}
\end{equation}

Let us revert back to the scenario of two Bobs. Corresponding to every pair of incompatible measurements $\{M_k\}_k$ and $\{N_l\}_l$, there exists at least one LO (which is a subset of LOCC) state discrimination task for which this upper bound can be achieved. Before discussing the actual scenario, let us first state the semi-definite program (SDP), through which ROI of a set of measurements, can be expressed.


The forms of the primal SDPs to determine the ROIs of the measurements $\{M_k\}_k$, and $\{N_l\}_l$ are given by
\begin{eqnarray}
1+I_M=\min_{s,\{\tilde{G}_\textbf{c}\}} s &&  \label{primal1}\\
\text{such that}~~~ &&  \sum_\textbf{c} D_\textbf{c}(c|k)\tilde{G}_\textbf{c}\geq M_{c|k}\nonumber\\
&& \sum_{\textbf{c}}\tilde{G}_\textbf{c}=s\mathbbm{1}\text{, } \tilde{G}_\textbf{c}\geq 0.\nonumber 
\end{eqnarray}
Here, $s=1+r$, where $r$ is defined in \eqref{sdp}. $\tilde{G}_\textbf{c} = sG_\textbf{c}$ and the positivity of $\Lambda_{c|k}$ in \eqref{sdp}, leads to the inequality $\sum_\textbf{c}D_\textbf{c}(c|k)\tilde{G}_\textbf{c}\geq M_{c|k}$, where $p(c|k, \lambda) = \sum_{\textbf{c}}D_{\textbf{c}}(c|k)p(\textbf{c}|\lambda)$, $\textbf{c} = \textbf{c}_1\textbf{c}_2\dots\textbf{c}_n$, a string of outcomes and $D_{\textbf{c}}(c|k) = \delta_{c,\textbf{c}_k}$. The $I_M$ defined in Eq.~\eqref{primal1} quantifies the incompatibility of the set of measurements available on Bob1's side. In a similar manner, the incompatibility of the set of measurements, $\{N_l\}_l$, accessible to Bob2 can also be defined. The corresponding SDP can be formulated as
\begin{eqnarray}
\text{and}~~1+I_N=\min_{t,\{\tilde{H}_\textbf{d}\}} t && \label{primal2}\\
\text{such that}~~~ &&  \sum_\textbf{d} E_\textbf{d}(d|l)\tilde{H}_\textbf{d}\geq N_{d|l}\nonumber\\
&& \sum_{\textbf{d}}\tilde{H}_\textbf{d}=t\mathbbm{1}\text{, } \tilde{H}_\textbf{d}\geq 0. \nonumber 
\end{eqnarray}

Mathematically, the parameters $s$ and $t$ carry the same meaning, with the only difference being that the optimal $s$ and $t$ are equal to unity added with ROI of measurements available on Bob1's side and Bob2's side, respectively.

The corresponding dual SDPs can be expressed as
\begin{eqnarray}
1+I_M=&&\max_{X,\{w_{ck}\}} \text{tr}\left[\sum_{c,k}w_{ck}M_{c|k}\right] \label{dual1}\\
\text{such that}~~~&&  X\geq \sum_{c,k}w_{ck}D_\textbf{c}(c|k),\nonumber\\
&& w_{ck}\geq 0\text{, } \text{tr}[X]=1,	\nonumber
\end{eqnarray}
\begin{eqnarray}
\text{and}~~1+I_N=&&\max_{Y,\{z_{dl}\}} \text{tr}\left[\sum_{d,l}z_{dl}N_{d|l}\right]\label{dual2}\\
\text{such that}~~~&&  Y\geq \sum_{d,l}z_{dl}E_\textbf{d}(d|l),\nonumber\\
&& z_{dl}\geq 0\text{, } \text{tr}[Y]=1,	\nonumber
\end{eqnarray}
where $w_{ck}$, $X$, $z_{dl}$, and $Y$ are the dual variables. See Ref.~\cite{cavalcanti1} for a more detailed treatment of these primal and dual problems.

We consider the dual variables $w_{ck}^*$, $X^*$, $z_{dl}^*$ and $Y^*$ for which the optimizations in Eqs.~\eqref{dual1} and~\eqref{dual2} are achieved and write
%
\begin{equation}
1+I_M=\text{tr}\left[\sum_{c,k}w^*_{ck}M_{c|k}\right]\text{ and }1+I_N=\text{tr}\left[\sum_{d,l}z^*_{dl}N_{d|l}\right]. \label{13}  
\end{equation}
Let us now introduce some new variables, given by
\begin{eqnarray}
&&M^*=\text{tr}\left[\sum_{c,k}w_{ck}^*\right]\text{, }N^*=\text{tr}\left[\sum_{d,l}z_{dl}^*\right]\text{, }\nonumber\\
&&q^*(cd,kl)= \frac{\text{tr}[w_{ck}^*]\text{tr}[z_{dl}^*]}{M^* N^*},\nonumber\\
\text{and}~~~&&\rho^*_{cd|kl}=\frac{w^*_{ck}\otimes z^*_{dl}}{\text{tr}[w_{ck}^*]\text{tr}[z_{dl}^*]}=\frac{w^*_{ck}\otimes z^*_{dl}}{M^*N^*q^*(cd,kl)}.\label{12}
\end{eqnarray}
The dual variables $w_{ck}^*$ and $z_{dl}^*$ are positive, hermitian operators. So, $\rho^*_{cd|kl}$ is a quantum state. We now 
state
the corresponding state discrimination task: Alice can choose an ensemble $\mathcal{E}^*_{kl}$ with probability $q^*(kl)$ which consists of bipartite states $\rho^*_{cd|kl}$. The probability of choosing a state $\rho^*_{cd|kl}$ from $\mathcal{E}_{kl}$ is $q^*(cd|kl) = \frac{q^*(cd,kl)}{q^*(kl)}$, ${q^*(kl)}=\sum_{cd} q^*(cd,kl)$. Then, Alice prepares a quantum system in the state $\rho^*_{cd|kl}$ and the subsystems are sent to Bob1 and Bob2. The task of Bob1 and Bob2 is to identify $cd$. 
To complete the task successfully, Bob1 and Bob2 choose measurements from the sets of measurements $\{M_k\}_k$ and $\{N_l\}_l$. Since in case of SD1, Bob1 and Bob2 know the ensemble from which Alice has chosen the state, prior to their measurements, they can choose the measurement based on the information of $kl$. Let us assume that Bob1 and Bob2 choose the measurements $M_{k'}$ and $N_{l'}$ with probabilities $p(k'|kl)$ and $p(l'|kl)$, respectively. But in case of SD2, information of $kl$ is considered to be unknown before the performance of measurements. Thus the measurements have to be chosen independent of the value of $kl$. We assume that for SD2, the measurements $M_{k'}$ and $N_{l'}$ are chosen with probabilities $p(k')$ and $p(l')$, respectively.

The operators associated with the measurements $M_{k'}$ and $N_{l'}$ are given by $\{M_{c'|k'}\}_{c'}$ and $\{N_{d'|l'}\}_{d'}$. For SD1, i.e., state discrimination with pre-measurement information we consider a specific strategy, i.e., $p(k'|kl)=\delta_{kk'}$, $p(l'|kl)=\delta_{ll'}$ and $p(cd|c',d',kl)=\delta_{cc'}\delta_{dd'}$. It can be proved that this state discrimination task achieves the bound. The proof has been presented Appendix~\ref{A2}. The state discrimination task can be generalized to $n$-parties and correspondingly the bound given in inequality \eqref{gen1} can also be proved to be achievable.

\section{The local bounds vs the global one}.
\label{sec5}
In Ref.~\cite{cavalcanti1}, the authors have considered a state discrimination task that is different from the ones considered until now, and where only two parties were involved, say Alice and Bob. In that protocol, Alice chose an ensemble $\mathcal{E}_y$ with probability $q(y)$. She then prepared a quantum system in a state $\rho_{b|y}$, taken from $\mathcal{E}_y$ with probability $q(b|y)$. After its preparation, she had sent the entire quantum system to Bob. She also informed Bob about the value of $y$. Bob's task was to identify $b$. Since in that situation, Bob was holding the complete state, i.e., was not sharing the state with any third party, he was able to do measurement on the entire system. Thus the restriction of local operations and classical communication was not applicable. But there also Bob was allowed to perform only a set of measurements, say $\{Q_x\}_x$. We remember that the suffix, $x$, written outside the second bracket of the expression $\{Q_x\}_x$ indicates the running variable which generates the set. At this point, depending on the information of $y$, Bob chose a particular measurement, $Q_x$ from the set of measurements, $\{Q_x\}_x$  with probability $p(x|y)$. The maximum PSG using the set of measurements $\{Q_x\}_x$ can be denoted by $\mathcal{P}^I(\{\mathcal{E}_y\},\{Q_x\})$. The maximum PSG optimized over the set of compatible measurements, when no information is available about $y$ until the measurement is performed, can be denoted as $\mathcal{P}^C(\{\mathcal{E}_y\})$. It was shown in Ref.~\cite{cavalcanti1} that 
\begin{equation}
 \frac{\mathcal{P}^I(\{\mathcal{E}_y\},\{Q_x\})}{\mathcal{P}^C(\{\mathcal{E}_y\})}\leq 1+I_Q,  \label{gbound} 
\end{equation}
where $I_Q$ is the ROI of $\{Q_x\}_x$. This is certainly a ``global" bound on the achievable advantage of incompatibility, because here the entire state is available to Bob for measurements. In this paper, we considered the state to be shared between two distant parties, Bob1 and Bob2, who were only allowed to do local operations and classical communication on their parts of the system. Thus we determined a ``local" bound on the achievable advantage of incompatibility. 
We now want to compare the global bound, expressed in \eqref{gbound},
with the local ones, obtained in \eqref{ublo} and \eqref{ublocc}.

Let incompatibility of the set of global measurements $\{M_k\otimes N_l\}$ be $I_{M\otimes N}$. It can be shown that $1+I_{M\otimes N}=(1+I_M)(1+I_N)$. [See Appendix~\ref{A3} for a proof.]
 Since LO is a subset of LOCC and LOCC is a subset of separable operations~\cite{paper1,LOCC3},
 we have $P^I_\text{LO\cancel{CC}}(\{\mathcal{E}_y\},\{M_k\},\{N_l\})\leq P^I_{\text{LOCC}}(\{\mathcal{E}_y\},\{M_k\},\{N_l\}) \leq \mathcal{P}^I(\{\mathcal{E}_y\},\{M_k\otimes N_l\})\leq (1+I_{M\otimes N}) \mathcal{P}^C(\{\mathcal{E}_y\},\{M_k\otimes N_l\})$, for any set of ensembles $\{\mathcal{E}_y\}$. On the other hand, ${P}^C_{\text{LO\cancel{CC}/LOCC}}(\{\mathcal{E}_y\},\{M_k\}, \{N_l\})\leq \mathcal{P}^C(\{\mathcal{E}_y\},\{M_k\otimes N_l\})$ is also true because there exist examples for which such an equality holds. For instance, let us consider that Alice has only one ensemble $\mathcal{E}_0$, i.e $q(y)=\delta_{y,0}$. The ensemble consists of equally probable two-qubit maximally entangled states. Since the states are orthogonal, they can be globally distinguished (by measuring onto the basis of the states).
 Thus we have $\mathcal{P}^C(\mathcal{E}_0)=1$. But even if classical communication is allowed, they can never be deterministically distinguished using LOCC \cite{LOCC7}. Thus $P^C(\mathcal{E}_0)<\mathcal{P}^C(\mathcal{E}_0)$. Hence, we can say the bounds given in \eqref{ublo} and \eqref{ublocc} restrict the PSG using incompatible measurements more than what would in general be possible via the previously known global bound.
 \section{Absence of nonlocality in optimal local state discrimination.} Since $1+I_{M\otimes N}=(1+I_M)(1+I_N)$, we see the bounds on the ratios of the probabilities, i.e. on $P_{\text{LO\cancel{CC}/LOCC}}^I/P_{\text{LO\cancel{CC}/LOCC}}^C$, given in \eqref{ublo} and \eqref{ublocc} are equal with the global bound on $\mathcal{P}^I/\mathcal{P}^C$ presented in \cite{cavalcanti1}. For each set of local measurements, there exists a corresponding state discrimination task where the bounds are achievable. This indicates that there is no ``nonlocality''  present in the ratio of the success probabilities using incompatibility 
 measures, in the optimal state discrimination process.  Here, ``nonlocality'' is being used in the sense of a difference between the ratios of the probabilities, in global and local state distinguishability. Note however that nonlocality in the individual probabilities might still be present which might have cancelled out at the time of taking the ratios.




\section{Conclusion}.\label{sec6}
Incompatibility of observables is a signature quantum mechanical property, which is active in arguably all quantum tasks. It was known 
that incompatibility can be used as a resource in  quantum state discrimination protocols. 

Behavior of shared systems is a widely-researched topic which offers various fascinating results. These can then be used to develop quantum technologies.  The difference between the ability to distinguish shared quantum states using global and local operations provides evidence of “nonlocality” present in the considered situation.

In this paper, we tried to forge a bridge 
between 
the efficiency of local quantum state discrimination using incompatible measurements and 
the relevant
quantum measurement incompatibility.
We have considered local quantum state discrimination tasks, where in one case, only local quantum operations were allowed, and in the other,  unidirectional classical communication was allowed along with the local operations. We have presented an upper bound on the ratio of the probability of successfully guessing the sent quantum state using incompatible measurements and the maximum probability of the same using any set of compatible ones. This upper bound is the same for both local operations, and local operations assisted by unidirectional classical communication, and is an achievable bound in at least one local quantum state discrimination exercise. We have compared the local bound with the existing global bound. We have shown that the optimal local quantum state discriminations do not reveal any nonlocality in the ratios of the probabilities between incompatible and compatible measurements.
 \section{Acknowledgements}.
We acknowledge partial support from the Department of Science and Technology, Government of India, through QuEST Grant No.~DST/ICPS/QUST/Theme-3/2019/120.

\appendix

\section{State discrimination using local operations and classical communication}\label{A1}

Let us consider the same couple of state discrimination tasks, i.e., state discrimination task with pre-measurement information (SD1) and with post-measurement
information (SD2) as described in the main text, the only difference being that Bob1 and Bob2 now have the facility of using classical communication between the local measurements.
The state discrimination task with local operations and classical communications (LOCC) considered in this regard is described as follows.



\emph{State discrimination using LOCC.}--If the parties are allowed to avail any sequence of classical communication between the local measurements in order to accomplish the given state discrimination task, it is called state discrimination by LOCC \cite{LOCC3.5, LOCC3, LOCC4, LOCC5, LOCC6, LOCC7, LOCC9, LOCC10, LOCC11, LOCC12, LOCC13, LOCC14, LOCC15, LOCC16, LOCC17, LOCC18, LOCC19, LOCC20, LOCC21, LOCC22, LOCC24, LOCC25, LOCC26, LOCC27, LOCC28, LOCC29, LOCC30, LOCC31, LOCC32, LOCC34, LOCC33, LOCC35, LOCC36, LOCC38, LOCC40, LOCC44, LOCC45, LOCC49, LOCC50, LOCC55, LOCC56, LOCC65, LOCC66, LOCC67, LOCC68}. Thus, in this case, one party can choose her/his measurement, based on the measurement outcomes of the other parties. 

 Nevertheless, only one round of measurements is considered here, i.e., Bob1 first chooses his measurement, $M_k$, with probability $p(k|y)$ (for SD1) or $p(k)$ (for SD2) and Bob2 chooses his measurement, $N_l$, after getting the knowledge of Bob1's measurement outcome (say, $M_{c|k}$ clicked), with probability $p(l|c,y)$ (for SD1) or $p(l|c)$ (for SD2). In this scenario, the the maximum PSG in case of SD1, using a particular set of measurements, $\{M_k\}$ and $\{N_l\}$, is given by
\begin{eqnarray}
P^I_{\text{LOCC}}(\{\mathcal{E}_y\},\{M_k\},\{N_l\}) = \max_{p(k|y),p(l|c,y), p(b|c,d,y)}\sum_{c,d,k,l,b,y}q(y)q(b|y)\nonumber \\
\tr[\rho_{b|y}M_{c|k}\otimes N_{d|l}]p(k|y)p(l|c,y)p(b|c,d,y).\nonumber
\end{eqnarray} 
In case of SD2, the maximum PSG using compatible measurements is 
\begin{eqnarray*}
P^C_{g,\text{LOCC}}(\{\mathcal{E}_y\},\{G_k\},\{H_l\}) = \max\sum_{c,d,k,l,b,y}q(y)q(b|y)
\\ \tr[\rho_{b|y}G_{c|k}\otimes H_{d|l}]p(k)p(l|c)p(b|c,d,y),
\end{eqnarray*}
where the maximization has to be taken over the set of parameters $\{\{G_k\},\{H_l\},p(k),p(l|c), p(b|c,d,y)\}$.

Suppose again that ROIs of the measurements $\{M_k\}_k$ and $\{N_l \}_l$ are $I_M$ and $I_N$. Now multiplying both side of the inequality (8) of the manuscript by $q(y)q(b|y)p(k|y)p(l|c,y)p(b|c,d,y)\rho_{b|y}$, taking trace, and summing over $c$, $d$, $k$, $l$, $b$, $y$, we get
\begin{widetext}
\begin{eqnarray}
\sum_{c,d,k,l,b,y} q(y)q(b|y)p(k|y)p(l|c,y)p(b|c,d,y)\tr[\rho_{b|y}M_{c|k}\otimes N_{d|l}]\leq (1+I_M)(1+I_N)\sum_{c,d,k,l,b,y,\lambda} q(y)q(b|y)p(k|y)p(l|c,y)p(b|c,d,y)\nonumber\\p^*(c|k,\lambda)p^*(d|l,\nu)\tr[\rho_{b|y}G^*_\lambda\otimes H^*_\nu].~~~~
\end{eqnarray} 
By defining a new probability distribution $p(b|\lambda,\nu, y)=\sum_{c,d,k,l}p(b|c,d,y)p(k|y)p(l|c,y)p^*(c|k,\lambda)p^*(d|l,\nu)$, we have
\begin{eqnarray}
\sum_{c,d,k,l,b,y} q(y)q(b|y)\tr[\rho_{b|y}M_{c|k}\otimes N_{d|l}]p(k|y)p(l|c,y)p(b|c,d,y)\leq (1+I_M)(1+I_N)\sum_{b,y,\lambda,\nu} q(y)q(b|y)\tr[\rho_{b|y}G^*_\lambda\otimes H^*_\nu]p(b|\lambda,\nu,y).~~~~
\end{eqnarray}
\end{widetext}
The expression, $\sum_{b,y,\lambda,\nu} q(y)q(b|y)\tr[\rho_{b|y}G^*_\lambda\otimes H^*_\nu]p(b|\lambda,\nu,y)$, in the RHS of the above inequality represents the success probability using a particular setting of local measurements. Thus, it is less than or equal to the maximum PSG using compatible local measurements, i.e.,
\begin{eqnarray}
&&\sum_{c,d,k,l,b,y} q(y)q(b|y)\tr[\rho_{b|y}M_{c|k}\otimes N_{d|l}]p(k|y)p(l|c,y)p(b|c,d,y)\nonumber\\&&\leq (1+I_M)(1+I_N)P_{\text{LOCC}}^C(\{ \mathcal{E}_y\}).
\end{eqnarray} 
Maximizing the LHS of the above equation over the probability distributions $p(k|y)$, $p(l|c,y)$, and $p(b|c,d,y)$, we have
\begin{eqnarray}
P^I_{\text{LOCC}}(\{ \mathcal{E}_y\}, \{ M_k\},\{ N_l\})&\leq& (1+I_M)(1+I_N)P_{\text{LOCC}}^C(\{ \mathcal{E}_y\}),\nonumber\\
\text{i.e., }\frac{P^I_{\text{LOCC}}(\{ \mathcal{E}_y\}, \{ M_k\}, \{N_l\})}{P_{\text{LOCC}}^C(\{ \mathcal{E}_y\})}&\leq& (1+I_M)(1+I_N)\label{ublocc}.
\end{eqnarray}
The RHS in the above bound is the same as in the bound given in inequality (12) of the manuscript. Thus, in the present setup, we 
do not have an improved
bound by using classical communication.\\

\textbf{More parties.}
The case of more parties here is again similar to that in the case of LO without CC described in the manuscript.


\section{Achievability of the Bound}
\label{A2}
Here we show that corresponding to every pair of incompatible measurements $\{M_k\}_k$ and $\{N_l\}_l$, there exists at least one state discrimination task with LO without CC for which this upper bound can be achieved. Before going into the actual proof, let us first state the semi-definite program (SDP), through which ROI of a set of measurements, can be expressed. The forms of the primal SDPs to determine the ROIs of the measurements $\{M_k\}_k$, and $\{N_l\}_l$ are given by
\begin{eqnarray}
1+I_M=\min_{s,\{\tilde{G}_\textbf{c}\}} s &&  \label{primal1}\\
\text{such that}~~~ &&  \sum_\textbf{c} D_\textbf{c}(c|k)\tilde{G}_\textbf{c}\geq M_{c|k},
 \sum_{\textbf{c}}\tilde{G}_\textbf{c}=s\mathbbm{1}\text{, } \tilde{G}_\textbf{c}\geq 0,\nonumber 
\end{eqnarray}
\begin{eqnarray}
\text{and}~~1+I_N=\min_{t,\{\tilde{H}_\textbf{d}\}} t && \label{primal2}\\
\text{such that}~~~ &&  \sum_\textbf{d} E_\textbf{d}(d|l)\tilde{H}_\textbf{d}\geq N_{d|l}
 \sum_{\textbf{d}}\tilde{H}_\textbf{d}=t\mathbbm{1}\text{, } \tilde{H}_\textbf{d}\geq 0. \nonumber 
\end{eqnarray}
Here, $s=1+r$, where $r$ is defined in (2) of the manuscript. $\tilde{G}_\textbf{c} = sG_\textbf{c}$ and the positivity of $\Lambda_{c|k}$ in (2), leads to the inequality $\sum_\textbf{c}D_\textbf{c}(c|k)\tilde{G}_\textbf{c}\geq M_{c|k}$, where $p(c|k, \lambda) = \sum_{\textbf{c}}D_{\textbf{c}}(c|k)p(\textbf{c}|\lambda)$, $\textbf{c} = \textbf{c}_1\textbf{c}_2\dots\textbf{c}_n$, a string of outcomes and $D_{\textbf{c}}(c|k) = \delta_{c,\textbf{c}_k}$. 
A similar situation is true for Eq.~(\ref{primal2}).
The corresponding dual SDPs can be expressed as
\begin{eqnarray}
1+I_M=&&\max_{X,\{w_{ck}\}} \text{tr}\left[\sum_{c,k}w_{ck}M_{c|k}\right] \label{dual1}\\
\text{such that}~~~&&  X\geq \sum_{c,k}w_{ck}D_\textbf{c}(c|k),w_{ck}\geq 0\text{, } \text{tr}[X]=1,	\nonumber
\end{eqnarray}
\begin{eqnarray}
\text{and}~~1+I_N=&&\max_{Y,\{z_{dl}\}} \text{tr}\left[\sum_{d,l}z_{dl}N_{d|l}\right]\label{dual2}\\
\text{such that}~~~&&  Y\geq \sum_{d,l}z_{dl}E_\textbf{d}(d|l), z_{dl}\geq 0\text{, } \text{tr}[Y]=1,	\nonumber
\end{eqnarray}
where $w_{ck}$, $X$, $z_{dl}$, and $Y$ are the dual variables. See Ref.~\cite{cavalcanti1} for a more detailed treatment of these primal and dual problems.

We consider the dual variables $w_{ck}^*$, $X^*$, $z_{dl}^*$ and $Y^*$ for which the optimizations in Eqs.~\eqref{dual1} and~\eqref{dual2} are achieved and write
%
\begin{equation}
1+I_M=\text{tr}\left[\sum_{c,k}w^*_{ck}M_{c|k}\right]\text{ and }1+I_N=\text{tr}\left[\sum_{d,l}z^*_{dl}N_{d|l}\right]. \label{13}  
\end{equation}
Let us now introduce some new variables, given by
\begin{eqnarray}
&&M^*=\text{tr}\left[\sum_{c,k}w_{ck}^*\right]\text{, }N^*=\text{tr}\left[\sum_{d,l}z_{dl}^*\right]\text{, }\nonumber\\&& q^*(cd,kl)= \frac{\text{tr}[w_{ck}^*]\text{tr}[z_{dl}^*]}{M^* N^*}\text{, }\text{and}~~~\rho^*_{cd|kl}=\frac{w^*_{ck}\otimes z^*_{dl}}{\text{tr}[w_{ck}^*]\text{tr}[z_{dl}^*]}.\nonumber
\end{eqnarray}
The dual variables $w_{ck}^*$ and $z_{dl}^*$ are positive, hermitian operators. So, $\rho^*_{cd|kl}$ is a quantum state. We now 
state
the corresponding state discrimination task: Alice can choose an ensemble $\mathcal{E}^*_{kl}$ with probability $q^*(kl)$ which consists of bipartite states $\rho^*_{cd|kl}$. The probability of choosing a state $\rho^*_{cd|kl}$ from $\mathcal{E}_{kl}$ is $q^*(cd|kl) = \frac{q^*(cd,kl)}{q^*(kl)}$, ${q^*(kl)}=\sum_{cd} q^*(cd,kl)$. Then, Alice prepares a quantum system in the state $\rho^*_{cd|kl}$ and the subsystems are sent to Bob1 and Bob2. The task of Bob1 and Bob2 is to identify $cd$. To complete the task successfully, Bob1 and Bob2 choose the measurements $M_{k'}$ and $N_{l'}$ from the sets of measurements $\{M_k\}_k$ and $\{N_l\}_l$ with probability $p(k'|kl)$ and $p(l'|kl)$ (in case of SD1) or with probability $p(k')$ and $p(l')$ (in case of SD2) respectively. The operators associated with the measurements $M_{k'}$ and $N_{l'}$ are given by $\{M_{c'|k'}\}_{c'}$ and $\{N_{d'|l'}\}_{d'}$. We first consider SD1 with a specific strategy, i.e., $p(k'|kl)=\delta_{kk'}$, $p(l'|kl)=\delta_{ll'}$ and $p(cd|c',d',kl)=\delta_{cc'}\delta_{dd'}$. The probability of guessing the state of the system correctly using this particular strategy, can be denoted by $\tilde{P}_{\text{LO\cancel{CC}}}^{\text{SD1}}(\{\mathcal{E}_{kl}^*\},\{M_k\},\{N_l\})$. The maximum PSG using the optimal strategy, $P^I_{\text{LO\cancel{CC}}}(\{\mathcal{E}_{kl}^*\},\{M_k\},\{N_l\})$, will not be less than $\tilde{P}_{\text{LO\cancel{CC}}}^{\text{SD1}}(\{\mathcal{E}_{kl}^*\},\{M_k\},\{N_l\})$, and therefore we can write
\begin{eqnarray}
&&P^I_{\text{LO\cancel{CC}}}(\{\mathcal{E}_{kl}^*\},\{M_k\},\{N_l\})\geq \tilde{P}_{\text{LO\cancel{CC}}}^{\text{SD1}}(\{\mathcal{E}_{kl}^*\},\{M_k\},\{N_l\})\nonumber\\=&&\sum_{c,d,k,l} q^*(kl)q^*(cd|kl)\text{tr}[\rho^*_{cd|kl}M_{c|k}\otimes N_{d|l}]\nonumber\\=&&\sum_{c,d,k,l} \frac{1}{M^*N^*}\text{tr}[w_{ck}^*\otimes z_{dl}^* M_{c|k}\otimes N_{d|l}]\nonumber\\
=&&\frac{1}{M^* N^*}(1+I_N)(1+I_M). \label{E19}
\end{eqnarray} 
The last line in the above equation is written using Eq.~\eqref{13}. However, the maximum PSG using compatible measurements without having pre-measurement information is given by
\begin{eqnarray}
P_{\text{LO\cancel{CC}}}^C(\{\mathcal{E}_{kl}^*\})=&&\max_{G_{k'},H_{l'},p}\sum_\tau q^*(kl)q^*(cd|kl)\tr[\rho^*_{cd|kl}G_{c'|k'}\otimes H_{d'|l'}]\nonumber\\&& p(k')p(l')p(cd|c',d',kl).\nonumber
\end{eqnarray}
Here the maximization is taken over the compatible measurements $\{G_{k'}\}$ and $\{H_{l'}\}$, and the probability distributions $p$ = $\{\{p(k')\}_{k'},\{p(l')\}_{l'},\{p(cd|c',d',kl)\}_{c',d'}\}$. The summation is taken over the set of variables $\tau=\{c,c',d,d',k,k',l,l'\}$. Let the above maximization be achieved for the set of measurements $\{G^*_{k'}\}$ and $\{H^*_{l'}\}$, and probabilities $p^*=\{\{p^*(k')\}_{k'},\{p^*(l')\}_{l'},\{p^*(cd|c',d',kl)\}_{c',d'}\}$. Then we have
\begin{eqnarray}
P_{\text{LO\cancel{CC}}}^C(\{\mathcal{E}_{kl}^*\})=&&\sum_\tau q^*(kl)q^*(cd|kl)\tr[\rho^*_{cd|kl}G^*_{c'|k'}\otimes H^*_{d'|l'}]\nonumber\\&& p^*(k')p^*(l')p^*(cd|c',d',kl).\label{eeq12}
\end{eqnarray}

We can define two other sets of ensembles, $E_1=\{\mathcal{E}_k\}_k$ and $E_2=\{\mathcal{E}_l\}_l$, where each of the ensembles have states, $\{\rho_{c|k}\}_c$ and $\{\rho_{d|l}\}_d$. Now Alice independently chooses the ensembles $\mathcal{E}_k$ and $\mathcal{E}_l$ with probabilities $q(k)$ and $q(l)$ from the sets $E_1$ and $E_2$ respectively. And then from these ensembles, she chooses the states $\rho_{c|k}$ and $\rho_{d|l}$ with probabilities $q(c|k)$ and $q(d|l)$ respectively. We define these states and the probabilities in the following way:
\begin{eqnarray}
&&q^*(c,k)= \frac{\text{tr}[w_{ck}^*]}{M^*}\text{, }q^*(d,l)= \frac{\text{tr}[z_{dl}^*]}{N^*},q^*(k)=\sum_c q^*(c,k)\text{, }\nonumber\\
&&q^*(l)=\sum_d q^*(d,l),q^*(c|k)=\frac{q^*(c,k)}{q^*(k)}\text{, }q^*(d|l)=\frac{q^*(d,l)}{q^*(l)},\nonumber\\
&&\rho^*_{c|k}=\frac{w^*_{ck}}{\text{tr}[w_{ck}^*]}=\frac{w^*_{ck}}{M^*q^*(c,k)}
\text{and}~~~\rho^*_{d|l}=\frac{z^*_{dl}}{\text{tr}[z_{dl}^*]}=\frac{z^*_{dl}}{N^*q^*(d,l)}.\nonumber
\end{eqnarray}
It is clearly visible that we have defined these new set of states in such a way that $\rho_{cd|kl}=\rho_{c|k}\otimes\rho_{d|l}$. The probabilities are also related: $q^*(cd,kl)=q^*(c,k)q^*(d,l)$, therefore $q^*(kl)=q^*(k)q^*(l)$ and $q^*(cd|kl)=q^*(c|k)q^*(d|l)$. Thus choosing $\rho_{cd|kl}$ from a chosen ensemble $\mathcal{E}_{kl}$ is equivalent of independently choosing $\rho_{c|k}$ and $\rho_{d|l}$ from ensembles $\mathcal{E}_k$ and $\mathcal{E}_l$. Let us consider a particular ensemble $\mathcal{E}_{kl}$ and particular measurements $M_{k'}$ and $N_{l'}$. If we measure $M_{k'}$ and $N_{l'}$ on a state $\rho_{cd|kl}$ then the probability of getting outcome $c'$ and $d'$ is denoted by $p(c',d'|cd,kl)$. Then from Bayes' theorem, we know that if the measurement outcomes are $c'$ and $d'$, then the probability that the state on which the measurement have been done is $\rho_{cd|kl}$ is given by
\begin{eqnarray*}
    p(cd|c',d',kl)&=&\frac{p(c',d'|cd,kl)q^*(cd,kl)}{\sum_{cd} p(c',d'|cd,kl)q^*(cd,kl)}\nonumber\\
    &=&\frac{\tr\left[\rho_{cd|kl}M_{c'|k'}\otimes N_{d'|l'}\right]q^*(c,k)q^*(d,l)}{\sum_{cd} \tr\left[\rho_{cd|kl}M_{c'|k'}\otimes N_{d'|l'}\right]q^*(c,k)q^*(d,l)}\nonumber\\
    &=&p(c|c',k)p(d|d',l).
\end{eqnarray*}
Thus, in Eq. \eqref{eeq12}, we can substitute $p^*(cd|c',d',kl)$ by $p^*(c|c',k)p^*(d|d',l)$. Let us now decompose $p^*(c|c',k)$ and $p^*(d|d',l)$ in terms of deterministic probabilities, i.e., $p^*(c|c',k)=\sum_{\textbf{c}}p^*(\textbf{c}|c')D_{\textbf{c}}(c|k)$ and $p^*(d|d',l)=\sum_{\textbf{d}}p^*(\textbf{d}|d')E_{\textbf{d}}(d|l)$. $\textbf{c}$ is defined after SDP \eqref{primal2}. \textbf{d} can also be defined in a similar way. Using these decompositions, we can write
\begin{eqnarray}
P_{\text{LO\cancel{CC}}}^C&&(\{\mathcal{E}_{kl}^*\})=\sum_{\tau,\textbf{c},\textbf{d}} q^*(kl)q^*(cd|kl)\tr[\rho^*_{cd|kl}G^*_{c'|k'}\otimes H^*_{d'|l'}]\nonumber\\&&p^*(k')p^*(l')p^*(\textbf{c}|c')D_{\textbf{c}}(c|k)p^*(\textbf{d}|d')E_{\textbf{d}}(d|l).\label{11}
\end{eqnarray}
From the dual SDP formulations, i.e., from \eqref{dual1} and \eqref{dual2}, we have
\begin{eqnarray}
X^*\geq\sum_{c,k} D_{\textbf{c}}(c|k)w^*_{ck} ~~\text{and}~~ Y^*\geq\sum_{d,l} E_{\textbf{d}}(d|l)z^*_{dl}.
\end{eqnarray}
Using these we can write
\begin{eqnarray}
 \tr[X^*\otimes Y^* G^*_{c'|k'}\otimes H^*_{d'|l'}]p^*(k')p^*(l')p^*(\textbf{c}|c')p^*(\textbf{d}|d')\nonumber\\
  \geq M^*N^* \sum_{c,d,k,l}  q^*(kl)q^*(cd|kl)\text{tr}\left[\rho^*_{cd|kl}G^*_{c|k'}\otimes H^*_{d|l'}\right]\nonumber\\ D_{\textbf{c}}(c|k)E_{\textbf{d}}(d|l)p^*(k')p^*(l')p^*(\textbf{c}|c')p^*(\textbf{d}|d').\nonumber
\end{eqnarray}
If we take the sum over $c'$, $d'$, $k'$, $l'$, $\textbf{c}$, and $\textbf{d}$ on both sides of the above inequality then, since $\tr[{X}]=\tr[{Y}]=1$, the left hand side will become equal to unity, and the right hand side will become equal to $M^*N^*P_{\text{LO\cancel{CC}}}^C(\{\mathcal{E}_{kl}\}))$ (see Eq. \eqref{11}).
Thus we get
\begin{eqnarray*}
1\geq M^*N^*P^C_{g,\text{LO\cancel{CC}}}(\{\mathcal{E}^*_{kl}\}), ~~\text{or,}~~\frac{1}{M^* N^*}\geq P^C_{g,\text{LO\cancel{CC}}}(\{\mathcal{E}^*_{kl}\}).
\end{eqnarray*}
Using inequality \eqref{E19}, along with the above inequality, we get
\begin{equation}
\frac{P^I_{\text{LO\cancel{CC}}}(\{\mathcal{E}_{kl}^* \},\{M_k\},\{N_l\})}{P^C_{g,\text{LO\cancel{CC}}}(\{\mathcal{E}_{kl}^*\})}\geq (1+I_M)(1+I_N).
\end{equation}
But the LHS of the above expression is upper bounded by $(1+I_M)(1+I_N)$, see Eq. (12) of manuscript. Therefore, we conclude
\begin{equation}
\frac{P^I_{\text{LO\cancel{CC}}}(\{\mathcal{E}_{kl}^* \},\{M_k\},\{N_l\})}{P^C_{g,\text{LO\cancel{CC}}}(\{\mathcal{E}_{kl}^*\})}=(1+I_M)(1+I_N).\label{equality}
\end{equation}

We want to emphasize  the fact that for a general set of ensembles, $\{\mathcal{E}_{y}\}_y$, along with a pair of sets of local measurements, say $\{M_k\}_k$ and $\{N_l\}_l$, the ratio of $P^I_{\text{LO\cancel{CC}}}(\{\mathcal{E}_{y} \},\{M_k\},\{N_l\})$ and $P^C_{g,\text{LO\cancel{CC}}}(\{\mathcal{E}_{y}\})$ is upper bounded by $(1+I_M)(1+I_N)$. Thus, while the equality in~\eqref{equality} is valid for at least one state discrimination task, it may not be true for an arbitrary state discrimination task. But what we have proved is that corresponding to every $\{M_k\}_k$ and $\{N_l\}_l$, there exists at least one  state discrimination task, 
for which it is possible to get the maximum advantage of using locally incompatible measurements.


\section{Relation between global and local incompatibility}\label{A3}
Let us suppose that for two given sets of measurements $\{M_k\}$ and $\{N_l\}$, the primal SDPs, defined in Eqs.~\eqref{primal1} and~\eqref{primal2}, are achieved by the variables $s^*$, $t^*$, $\{\tilde{G}^*_{\boldsymbol{c}}\}$, and $\{\tilde{H}^*_{\boldsymbol{d}}\}$. Thus we have
\begin{eqnarray}
&&(1+I_M)(1+I_N)=s^*t^*,\label{eqref2}\\
&&\sum_\textbf{c,d} D_\textbf{c}(c|k)E_\textbf{d}(d|l)\tilde{G}^*_\textbf{c}\otimes\tilde{H}^*_\textbf{d}\geq M_{c|k}\otimes N_{d|l},\label{eqref3}\\
&&\sum_{\textbf{c,d}}\tilde{G}_\textbf{c}\otimes\tilde{H}_\textbf{d}=s^*t^*\mathbbm{1}
\text{, and }\tilde{G}_\textbf{c}\otimes \tilde{H}_\textbf{d}\geq 0. \label{eqref4}
\end{eqnarray}

Incompatibility of the global measurement $\{M_k\otimes N_l\}$ can be derived using the following primal SDP:
\begin{eqnarray}
1+I_{M\otimes N}=\min_{u,{J_{\boldsymbol{cd}}}} u,&& \label{primal3}\\
 \text{s.t. }&&\sum_{\boldsymbol{cd}} F_{\boldsymbol{cd}}(cd|kl)J_{\boldsymbol{cd}}\geq M_{c|k}\otimes N_{d|l},\nonumber\\
&& \sum_{\boldsymbol{cd}}J_{\boldsymbol{cd}}=u\mathbbm{1}\text{, }J_{\boldsymbol{cd}}\geq 0.\nonumber
\end{eqnarray}
From Eqs.~\eqref{eqref2},~\eqref{eqref3}, and~\eqref{eqref4}, we see that $J_{\boldsymbol{cd}}=\tilde{G}_{\boldsymbol{c}}\otimes \tilde{H}_{\boldsymbol{d}}$ and $u=s^*t^*$ is a feasible solution of the primal SDP defined in~\eqref{primal3}. Thus we have $1+I_{M\otimes N}\leq(1+I_M)(1+I_N)$. Using the dual form of SDP of $I_M$, $I_N$, and $I_{M\otimes N}$, and following the same path (algebra),
it can also be shown that $1+I_{M\otimes N}\geq(1+I_M)(1+I_N)$. Hence, we conclude that $$1+I_{M\otimes N}=(1+I_M)(1+I_N).$$

  
\end{document}